\documentclass{linear}  

\let\trimmarks\relax % remove WS markings

\usepackage{makeidx}
\usepackage{dewsb}    %%% Specific style file for this chapter

\makeindex

\begin{document}

\setcounter{chapter}{4}
%%%%%%%%%%%%%%%%%%%%%%%%%%%%%%%%%%%%%%%%%%%%%%%%%%%%%%%%%%%%%%%%%%%%%%%%%%%
%% Chapter: dynamical symmetry breaking
%%%%%%%%%%%%%%%%%%%%%%%%%%%%%%%%%%%%%%%%%%%%%%%%%%%%%%%%%%%%%%%%%%%%%%%%%%%

% Chapter references
\newcommand{\chHaber}{Ref.~\protect\refcite{Haber}}
\newcommand{\chFeng}{Ref.~\protect\refcite{Feng}}
\newcommand{\chMoenig}{Ref.~\protect\refcite{Moenig}}
\newcommand{\chHewett}{Ref.~\protect\refcite{Hewett}}

\chapter*{DYNAMICAL ELECTROWEAK SYMMETRY BREAKING}
\hfill
\begin{picture}(0,0)
\unitlength1mm
\put(-40,60){%
\parbox{4cm}{%
\begin{flushright}
  DESY 03--026\\
  TTP--03--09\\
  hep-ph/0303015
\end{flushright}}}
\end{picture}
\markboth{W.\ Kilian}{Dynamical electroweak symmetry breaking}
\thispagestyle{empty}

\vspace*{-2\baselineskip}
\author{Wolfgang Kilian}

\address{Deutsches Elektronen-Synchrotron DESY, 22603 Hamburg, Germany\\
E-mail: wolfgang.kilian@desy.de}
%% \address{Institut f\"ur Theoretische Teilchenphysik, University of Karlsruhe\\
%% 76128 Karlsruhe, Germany\\
%% E-mail: kilian@particle.uni-karlsruhe.de}

\begin{abstract}
Dynamical symmetry breaking provides a possible solution to the
electroweak hierarchy problem.  It requires new strong interactions
that are effective at some high-energy scale.  If there is no light
Higgs boson, this scale is constrained to be in the $\TeV$ range, and
signals of the new interactions can be observed, directly or
indirectly, in collider experiments.  Even if no observable states in
the Higgs sector are kinematically accessible, a Linear Collider will
cover the low-energy parameter space that arises in a systematic
model-independent analysis of dynamical electroweak symmetry breaking.

\vspace{\baselineskip}\noindent
To appear in: \emph{Linear Collider Physics in the New Millennium}
(K.~Fujii, D.~Miller and A. Soni, eds.), World Scientific.
\end{abstract}

\clearpage
\tableofcontents  

\clearpage
\section{Introduction}

\subsection{Particle masses}

The most striking property of the particle spectrum is clearly the
huge range in the fundamental scales.  On a logarithmic mass
scale, the known elementary particles (except for the
\index{neutrinos}neutrinos) populate a small region centered almost 20
orders of magnitude below the \index{Planck mass}Planck mass, the
fundamental scale of space-time geometry:
\vspace{\baselineskip}
\begin{center}
  \scalebox{0.8}{\includegraphics{dewsb-plots.1}}
\end{center}
\vspace{\baselineskip}
Their interactions, as far as we know, respect the gauge symmetry
principle.  Gauge invariance of the electroweak interactions in
particular forbids mass terms for {all} known elementary particles
(except for right-handed neutrinos).  In reality, all particles but
the photon are massive, so the \index{electroweak
symmetry breaking}electroweak $SU(2)_L\times U(1)_Y$ gauge symmetry is
softly broken.  It is manifest in the interactions, but hidden in the
mass spectrum.

The masses of vector bosons are best understood as mixing terms.  The
electroweak gauge bosons which form a $SU(2)_L$ triplet $W^{\pm0}$ and
a $SU(2)_L$ singlet $B$, are coupled to a multiplet of
\index{scalars, representation}\emph{scalar} states in a different
representation of the electroweak symmetry group.  Diagonalizing the
mass matrix, the resulting eigenstates consist of the massless
photon and the three massive vector bosons $W^{\pm}$
and~$Z$.  Their \index{vector bosons, longitudinal}longitudinal
components can be identified with three components of the scalar
multiplet.  Similarly, the \index{fermions, masses}(Dirac) masses and
mixings of the fermions originate from bilinear couplings of
left-handed and right-handed two-component (Weyl) fermions, which are
otherwise unrelated and belong to different $SU(2)_L\times U(1)_Y$
representations.

The three scalar states (i.e., the longitudinal vector bosons) in the
observed spectrum do not make up a complete linear representation of
$SU(2)_L$.  Therefore, a field theory based just on the degrees of
freedom that have been established experimentally is
non-renormalizable.  We can make use of such an effective theory to
get a consistent \index{low-energy expansion}low-energy expansion of
the relevant physics\cite{EFT}.  This formalism will be described
below in Sec.~\ref{sec:eft}\@.  However, at high energies a
non-renormalizable theory has an inherent \index{cutoff scale}cutoff
scale where it ceases to be predictive.  For instance, the scalars may
turn out as
\index{compositeness scale}composite at the cutoff scale, or there may be
additional scalars which have not yet been observed.  In any case, the
pattern of mass generation via mixing suggests a new interaction, and
the complexity of the flavor couplings indicates
a similar complexity of this new
\index{Higgs sector}\emph{Higgs} sector\cite{Higgs}, of which we
might just be scratching the surface.  In the \index{Standard
Model}Standard Model and its extensions the problem is solved by
\emph{brute force}, postulating the existence of scalar multiplets
with just the appropriate quantum numbers and couplings\cite{GSW},
but this need not be the solution chosen by Nature.

Since the actual properties of the Higgs sector are unknown, in a
first approach one may identify a parameter
\index{$v$ (electroweak scale)}\index{electroweak scale}$v$ as the characteristic scale of
electroweak symmetry breaking (EWSB), by convention taken as
\begin{equation}\label{v}\index{$G_F$ (Fermi constant)}
  v = (\sqrt2\,G_F)^{-1/2} \approx \q{246}{\GeV}.
\end{equation}
$v$ is an abstract quantity at this stage, and while its value is
fixed by the measurement of a low-energy process (muon decay), its
exact meaning depends on the underlying dynamics which is not yet
accessible directly.

\subsection{Exponentials}

\index{hierarchy problem}\index{electroweak symmetry breaking}
In searching for possible sources of the terms that softly break the
electroweak symmetry in the fundamental Lagrangian, one may take
either one of two different views:
\begin{enumerate}
\item
The electroweak gauge symmetry is an accidental approximate symmetry.
\item
The electroweak gauge symmetry is exact in the dynamics, but
spontaneously broken in the low-energy states.
\end{enumerate}
The first explanation is generally rejected since it implies that the
smallness of the breaking terms \index{Planck mass}($v/\MPl\sim
10^{-17}$) is pure coincidence.  For a natural explanation of this
ratio\cite{Su84}, we not only need to adopt the second scenario, but
we also are led to postulate \index{dynamical symmetry
breaking}\emph{dynamical} symmetry breaking.  Such an extremely small
number should be the result of solving fundamental dynamical equations
\emph{without} particularly small parameters.  Ultimately, we are
looking for a fundamental theory without adjustable parameters at all.

Fortunately, we know at least one solution to this problem.
Strongly-coupled quantum field theories generate large scale ratios
from the quantum effect of \index{renormalization
group}renormalization group running\cite{asf}.  They allow the
interpretation of the scale hierarchy as an exponential factor
\begin{equation}\label{exponential}
  \MPl/v = e^{(\text{number of order 10})}
\end{equation}
{within} the context of quantum field theory.\footnote{Alternatively,
such an exponential could be a consequence of the space-time
structure.  This possibility is discussed in \chHewett.}

\index{QCD}QCD-type condensation is the one example where this
actually happens in particle physics.  Neglecting all other
interactions for simplicity, we could solve the renormalization group
equation for the running strong coupling constant $g_s(\mu)$ with a
``natural'' initial condition at the \index{Planck mass}Planck scale,
say $g_s(\MPl)=1/2$.  Identifying the QCD scale $\Lambda_{\rm QCD}$
with the scale where $g_s(\mu)$ becomes strong at low energies, we
obtain a scale ratio $\MPl/\Lambda_{\rm QCD}$ similar to
(\ref{exponential}).  The exponent is proportional to the inverse of
the initial value of $\alpha_s=g_s^2/4\pi$ at the
high scale.  In the QCD case, the low-energy singularity in the
perturbative evolution of the coupling is resolved by the condensation
of gluons and of left-handed and right-handed fermions, breaking the
chiral symmetry at the scale $\Lambda_{\rm QCD}\lesssim\q{1}{\GeV}$.
Incidentally, this mechanism would also trigger \index{electroweak
symmetry breaking}EWSB in the $\GeV$ range, if the electroweak
symmetry was not already broken at the higher scale $v=\q{246}{\GeV}$.

In fact, \index{dynamical symmetry breaking}dynamical symmetry
breaking is the explanation for a wide variety of physical phenomena,
from \index{superconductivity}superconductivity to the
laser effect.  The overall description is simple.  Out of
the fundamental fields of the model one constructs scalar field
multiplets $\Phi$ with nonvanishing quantum numbers under the symmetry
in question.  If the effective potential for such a field has a
nontrivial solution which is energetically favorable, it will get a
\index{vacuum expectation value} vacuum expectation value, breaking
the symmetry and allowing for new effective couplings which exact
symmetry would forbid.  There is no need for $\Phi$ to correspond to
independent observable degrees of freedom, since it can be a composite
of other fields present in the theory (such as Cooper pairs in BCS
superconductivity).  Thus, while fundamental \index{Higgs boson}Higgs
particles as $\Phi$ quanta are not excluded, they are not necessary
for the Higgs mechanism of EWSB to work.

\subsection{Higgs or no Higgs?}

\index{Higgs boson}
Just as in QCD, one can associate a scale $\Lambda$ with dynamical
symmetry breaking, the
\index{compositeness scale}\emph{compositeness scale}.  $\Lambda$ might be
significantly higher than the EWSB scale \index{$v$ (electroweak
scale)}\index{electroweak scale}$v$.  Then, in the intermediate scale
range $v\ldots\Lambda$ the model will reduce to a renormalizable
effective theory.  This can only be the case if the scalars which
provide the longitudinal $W$ and $Z$ states are accompanied by extra
scalar states such that a complete \index{scalars,
representation}linear representation of $SU(2)_L\times U(1)_Y$ is
formed.  The extra states can be observed as particles, the Higgs
bosons (cf.\ \chHaber).

As will be discussed below, in models without an observable Higgs
state the compositeness scale cannot be higher than\cite{NDA}
\begin{equation}\label{Lambda-TC}
  \Lambda \lesssim 4\pi v \approx \q{3}{\TeV}.
\end{equation}
If there is a single Higgs boson with mass $m_H$, the compositeness
scale is constrained instead by
\begin{equation}\label{Lambda-prime}
  \Lambda' \lesssim m_H \exp{\frac{\Lambda^2}{12 m_H^2}} \qquad
  \text{with $\Lambda=4\pi v$},
\end{equation}
the location of the \index{Landau pole}Landau pole of the
Higgs self-coupling.  The relation (\ref{Lambda-prime}) implies that
the concept of a Higgs boson makes sense only if
$m_H\lesssim\q{1}{\TeV}$, such that $\Lambda'>m_H$\cite{Hlimit}.

The ratio $\Lambda'/m_H$ could be large, but dynamical symmetry
breaking by itself provides no obvious mechanism for this.  One would
rather expect this ratio to be of order one, so the window for
composite Higgs states is narrow.  On the other hand, scalar states
could actually be elementary degrees of freedom.  This is consistent
with a dynamical solution of the \index{hierarchy problem}hierarchy
problem only if their masses are protected by a symmetry.  In that
case, the scalar interactions which trigger EWSB could only be
generated indirectly, which shifts dynamical symmetry breaking to a
new (hidden) sector of the theory.  This is the way
\index{supersymmetry}supersymmetric models are constructed, which are
considered in \chFeng.

Another possibility which opens some window for composite Higgs bosons
is the identification of the Higgs multiplet with
\index{pseudo-Goldstone bosons}pseudo-Goldstone
scalars generated by spontaneous symmetry breaking at a higher
scale\cite{tH79}.  Recently, realistic models with this structure
have been proposed, the so-called \index{Little Higgs
models}\emph{Little Higgs} models\cite{LHM}.  Like supersymmetric
models, they leave room for the electroweak scale being generated
dynamically through strong interactions, but in these scenarios the
effective theory is weakly interacting up to energies significantly
beyond the $\TeV$ scale.

By contrast, in the absence of light Higgs multiplets no weakly
interacting effective theory can be constructed that is valid up to
that energy range.  Instead, the strong interactions which accompany
dynamical scale generation may be directly coupled to Standard Model
particles and show up in scattering processes once the available
energies are sufficiently close to the new compositeness scale.
Therefore, one can not just expect to observe the effects
\emph{associated} with EWSB at colliders (e.g., Higgs particles,
superpartners, pseudo-Goldstone bosons), but there is hope to get a
handle on those (strongly-interacting) new degrees of freedom which
are \emph{responsible} for EWSB.

\subsection{Models of dynamical symmetry breaking}

\index{dynamical symmetry breaking}

Within the context of four-dimensional field theory, dynamical
symmetry breaking must be assigned to new gauge interactions.  This
is a common feature of all dynamical models of EWSB\cite{HS02}.  The
models differ in the role of fermions:
\begin{enumerate}
\item
Since, as in QCD, fermions which feel strong gauge interactions are
confined, it is likely that the fermions (\emph{technifermions})
directly associated with EWSB are unobservable at low energies.  This
is the original \index{technicolor}\emph{technicolor} (TC)
idea\cite{dynEWSB}.  Technifermion condensation is able to account
for effective scalar states which provide the longitudinal components
of $W$ and $Z$ bosons.
\item
Pure TC will not generate any left-right couplings for the observable
fermions, just as QCD generates constituent masses for quarks, but not
for leptons.  However, such couplings could be due to additional
dynamically broken gauge interactions at even higher energies which
are felt both by ordinary and by technifermions.  This mechanism is
known as \index{extended technicolor}\emph{extended technicolor}
(ETC)\cite{ETC}.  Below the ETC scale, such interactions lead to
\index{four-fermion couplings}four-fermion
couplings.  When technifermions condense at the
\index{compositeness scale}compositeness scale, the desired bilinear
couplings are generated. 
\item
Exchanging the roles of TC and ETC, a dynamically broken gauge
interaction might trigger EWSB by fermion condensation not too far
above the electroweak scale.  In that case, the affected fermions need
not be confined. The heavy \index{top quark}top quark is the prime
candidate \index{topcolor}(\emph{topcolor}) for such a
strongly-interacting object\cite{topcolor}.  Since for this mechanism
to actually work the top quark is somewhat too light, more recent
models that follow this pattern implement a combined
\index{topcolor-assisted technicolor}\emph{topcolor-assisted
technicolor} scheme\cite{tc2}.  Other models of this type involve the
condensation of \index{neutrinos}neutrinos, which might have large
Yukawa couplings despite their tiny physical masses and thus could
feel new strong interactions\cite{neu-con}.
\item
Alternatively, the physical top mass could be suppressed by a
two-state mixing effect \index{top see-saw model}(\emph{top
see-saw})\cite{t-ss}.  In such models, the electroweak scale $v$ is
typically suppressed compared to the compositeness scale $\Lambda$,
and there is room for \index{compositeness scale}\index{Higgs
boson}composite Higgs bosons in the intermediate range.
\end{enumerate}

The complicated pattern of flavor physics stands as the main obstacle
for constructing a simple theory of dynamical symmetry breaking.  It
is difficult to simultaneously accomodate (i) very light leptons and
quarks, (ii) a heavy top quark, (iii) the smallness of
\index{flavor-changing neutral currents}flavor-changing neutral
currents\cite{HS02}.  Turning the argument around, the fact that
\index{fermions, masses}fermion mass generation cannot be separated
from electroweak physics in strongly-interacting models, opens the
possibility that some of the puzzles of flavor physics are resolved at
future collider experiments.

\section{Effective Theories of Electroweak Interactions}
\label{sec:eft}

\subsection{The bottom-up approach}

\index{bottom-up approach}\index{effective Lagrangian}
While it is clearly worthwhile to search for signals of specific
models in collider experiments, the limited energy range of a
next-generation Linear Collider may not allow to access new states
associated with the Higgs sector directly.  In such a situation, a
model-independent treatment of the dynamics is more appropriate.
Fortunately, the formalism of effective (or phenomenological)
Lagrangians\cite{EFT} provides a generic framework for the bottom-up
description of electroweak interactions in the absence of a complete
renormalizable field theory.

At energies much below the $W$ and $Z$ (and Higgs) masses, the
dominant interactions of leptons and quarks are QED and
QCD interactions, governed by an effective Lagrangian of
\index{operators, dimension-2}\index{operators, dimension-3}\index{operators, dimension-4}operator dimension four and less:
\begin{equation}
\begin{split}\label{L3f}
  \LL &= \LL_3 + \LL_4 \\ &= -(\bar Q_L M_Q Q_R + \bar L_L M_L L_R + \hc)
  -(\bar N_L^c M_{N_L}N_L + \bar N_R^c M_{N_R}N_R) \\ &\quad + \bar
  Q_L i\fmslash D Q_L + \bar Q_R i\fmslash D Q_R + \bar L_L i\fmslash
  D L_L + \bar L_R i\fmslash D L_R - \tfrac14 A_{\mu\nu} A^{\mu\nu}
\end{split}
\end{equation}

For simplicity, we ignore \index{QCD}QCD interactions here.  The
building blocks consist, first of all, of left- and right-handed
\index{fermions, representation}quark and lepton fields,
\begin{equation}
  Q_L = \begin{pmatrix} U_L \\ D_L  \end{pmatrix},
  \quad
  Q_R = \begin{pmatrix} U_R \\ D_R  \end{pmatrix},
  \quad
  L_L = \begin{pmatrix} N_L \\ E_L  \end{pmatrix},
  \quad
  L_R = \begin{pmatrix} N_R \\ E_R  \end{pmatrix}.
\end{equation}
We omit generation and color indices, so all coupling constants should
be understood as matrices.  While gauge couplings are diagonal by
definition, the bilinear quark and lepton couplings $M_Q$ and $M_L$
and the \index{neutrinos}\index{Majorana masses}Majorana mass
matrices $M_{N_L}$ and $M_{N_R}$ are not.\footnote{Electroweak
symmetry requires $M_{N_L}$ to vanish, but $M_{N_L}$ is re-introduced
in the low-energy effective theory if $M_{N_R}$ is very large and the
right-handed neutrinos are integrated out.}  Upon diagonalization,
from this structure one obtains the fermion masses together with the
$3\times 3$ mixing matrices which determine the weak interactions of
quarks and neutrinos.

\index{QED}
QED gauge interaction are present in the dimension-four
terms which contain the electromagnetic field strength and covariant
derivative
\begin{equation}
  A_{\mu\nu} = \partial_\mu A_\nu - \partial_\nu A_\mu,
  \qquad
  D_\mu = \partial_\mu + ieq A_\mu,
\end{equation}
where $e$ is the positron charge and $q$ the multiplying factor for
the charge of a given fermion species.  In the doublet notation used
here, the normalized charge $q$ is a diagonal $2\times 2$ matrix which
reads
\begin{equation}
  q_{Q,L} = \tfrac12\left({y_{Q,L}+\tau^3}\right)
  \quad\text{with}\quad
  y_Q = \tfrac13 \quad\text{and}\quad
  y_L = -1.
\end{equation}
for the quarks and leptons, respectively.

At very low energies, the full electroweak symmetry is present only in
the higher-dimensional operators which induce weak interactions.
While magnetic-moment type operators
\index{operators, dimension-5}(dimension five)
\begin{equation}
  \LL_5 = \bar Q_R \mu_Q\sigma_{\mu\nu} A^{\mu\nu} Q_L 
        + \bar L_R \mu_L\sigma_{\mu\nu} A^{\mu\nu} L_L + \hc
\end{equation}
are strongly suppressed, \index{four-fermion
couplings}\index{operators, dimension-6}four-fermion operators
(dimension six)
\begin{equation}\label{L6-gen}
\begin{split}
  \LL_6 &=
   \sum_{f=Q_L,Q_R,L_L,L_R}
     s_{ijkl}(\bar f_i f_j)(\bar f_k f_l)
   + v_{ijkl}(\bar f_i\gamma^\mu f_j)(f_k\gamma_\mu f_l)
\end{split}
\end{equation}
are more significant.  Experiment has shown that their structure is
consistent with the specific factorizable form of the \index{Fermi
model}Fermi model,
\begin{equation}\label{L6-weak}
  \LL_6 = -4\sqrt2\,G_F \left(2J_\mu^+ J^{\mu -} 
                             + c_w^2 J_\mu^0 J^{\mu 0}\right)
\end{equation}
with the charged and neutral currents\index{weak currents}
\begin{align}
  J_\mu^\pm &= \tfrac{1}{\sqrt2}\left[
    \bar Q_L\tau^\pm\gamma_\mu Q_L 
  + \bar L_L\tau^\pm\gamma_\mu L_L
    \right]
  \\
  \begin{split}
  J_\mu^0 &= \tfrac{1}{c_w}\left[
    \bar Q_L\left(-q_Q s_w^2 + \tfrac{\tau^3}{2}\right)\gamma_\mu Q_L 
     + \bar Q_R(-q_Q s_w^2)\gamma_\mu Q_R \right.
    \\ &\qquad\left.+
    \bar L_L\left(-q_L s_w^2+\tfrac{\tau^3}{2}\right)\gamma_\mu L_L 
     + \bar L_R(-q_L s_w^2)\gamma_\mu L_R
    \right]
  \end{split}
\end{align}
Here, $s_w\approx 0.48$ is the sine of the weak mixing angle, and
$c_w=\sqrt{1-s_w^2}$.

We do not know with certainty that the factorizable structure of
four-fermion interactions is exact, but flavor-physics experiments
have not yet revealed any deviations from this picture.  Assuming that
deviations are absent, the \index{Fermi model}Fermi Lagrangian can be
rewritten with the help of the vector fields $W^\pm$ and $Z$,
\begin{equation}\label{L6-aux}
  \LL_6 = - g_W(W^{+\mu}J^+_\mu + W^{-\mu}J^-_\mu)  - g_Z(Z^\mu J^0_\mu)
        + \LL_{2(W)}
\end{equation}
where
\begin{equation}\label{L2W}
  \LL_{2(W)} = M_W^2 W^{+\mu}W^-_\mu + \tfrac12 M_Z^2 Z^\mu Z_\mu,
\end{equation}
such that (\ref{L6-aux}) becomes equivalent to (\ref{L6-weak}) when
the $W$ and $Z$ fields are integrated out.  After rearranging the basis,
\begin{align}
  \label{W-rot}
  W^+ &= \tfrac{1}{\sqrt2}(W^1-iW^2), &
  Z &= c_w W^3 - s_w B,\\
  W^- &=  \tfrac{1}{\sqrt2}(W^1+iW^2),  \label{ZA-rot} &
  A &= s_w W^3 + c_w B,
\end{align}
the Fermi Lagrangian assumes the form
\begin{equation}\label{L-fermi-gauge}
\begin{split}
  \LL &= \bar Q_L i{\fmslash D}_L Q_L + \bar Q_R i{\fmslash D}_R Q_R
  + \bar L_L i{\fmslash D}_L L_L + \bar L_R i{\fmslash D}_R L_R
  \\
  &\quad - \tfrac14 A_{\mu\nu} A^{\mu\nu} + \LL_{2(W)} + \LL_{3}
\end{split}
\end{equation}
with a local $SU(2)_L\times U(1)_Y$ symmetry manifest in the terms of
the first line.  With respect to this symmetry, the vector fields have
the usual gauge transformation properties, and the fermions couple via
\index{covariant derivative}covariant derivatives
\begin{align}
  \label{DL-new}
  D_{L\mu} &= \partial_\mu - ig' (q+\tfrac{\tau^3}{2}) B_\mu
                           + ig \tfrac{\tau^a}{2}W^a_\mu
  \\
  \label{DR-new}
  D_{R\mu} &= \partial_\mu - ig' q B_\mu.
\end{align}

To describe physics at and above the mass scale of the electroweak
vector bosons $W^\pm$ and $Z$, we have to include \index{operators,
dimension-4}kinetic terms
\begin{equation}\label{L4W}
  \LL_{4(W)} = -\tfrac12 \tr{\vW_{\mu\nu} \vW^{\mu\nu}}
              -\tfrac12 \tr{\vB_{\mu\nu} \vB^{\mu\nu}},
\end{equation}
where the \index{field strength tensors}field strength tensors are
defined in terms of vector fields $W^a$ ($a=1,2,3$) and $B$
\begin{align}
  \label{vW}
  \vW_{\mu\nu} &= \partial_\mu \vW_\nu  - \partial_\nu \vW_\mu
               + ig[\vW_\mu, \vW_\nu]
  \\
  \label{vB}
  \vB_{\mu\nu} &= \partial_\mu \vB_\nu  - \partial_\nu \vB_\mu
\end{align}
with $\vW_\mu = W_\mu^a\tfrac{\tau^a}{2}$ {and} $\vB_\mu = B_\mu
\tfrac{\tau^3}{2}$.  (This is not the most
general form allowed by the spontaneously broken gauge invariance, but
we postpone the discussion of anomalous couplings to
Sec.~\ref{sec:anomalous}.)

While the dimension-four part of the effective Lagrangian exhibits
full electroweak gauge symmetry, this symmetry is not manifestly
present in the dimension-two and dimension-three operators, the
fermion and gauge boson mass terms.  However, by introducing an extra
field \index{$\Sigma$ field}$\Sigma$ with a suitable transformation law,
this problem can formally be solved without losing the universality of
the effective-theory formalism\cite{CCWZ,EWChPT,ChPT}.  This field
parameterizes our ignorance about the true nature of the \index{Higgs
sector}Higgs sector.  While the actual dynamics at high energies could
be very complicated, at low energies a generic description is dictated
by symmetry considerations only.

The field $\Sigma(x)$ is a $2\times 2$ matrix which is defined to have
the appropriate behavior under gauge transformations $U(x)\in SU(2)_L$
and $V(x)=\exp(i\beta\tau^3)\in U(1)_Y$:
\begin{equation}\label{Sigma-trafo}
  \Sigma(x) \to U(x)\,\Sigma(x)\,V^\dagger(x).
\end{equation}
Spontaneous symmetry breaking is implemented by a nonzero
\index{vacuum expectation value}vacuum
expectation value
\begin{equation}\label{vev-sigma2}
  \vev{\tfrac12\tr{\Sigma^\dagger(x)\Sigma(x)}} = 1
\end{equation}
which for practical purposes can be replaced by
\begin{equation}
  \vev{\Sigma(x)} \to 1 \quad\text{for $x\to\infty$}.
\end{equation}
For perturbative calculations, it is often useful to adopt the
\index{unitary gauge}\emph{unitary} or \emph{unitarity} gauge where
$\Sigma(x) \equiv 1$, but when discussing the electroweak symmetry
structure, we should not impose this restriction.

A unitary matrix has three degrees of freedom, therefore the minimal
number of degrees of freedom parameterizing $\Sigma$ is three.  A
possible, but not unique, parameterization is given
by\cite{CCWZ,EWChPT,ChPT}
\begin{equation}\label{exp-par}\index{$\Sigma$ field}
  \index{scalars, representation}
  \Sigma(x) = \exp\left( -\tfrac{i}{v}\vw(x)\right)
  \quad\text{with $\vw(x)=w^a(x)\,\tau^a$; $a=1,2,3$.}
\end{equation}
where \index{$v$ (electroweak scale)}\index{electroweak scale}$v$ is the electroweak scale
(\ref{v}).  We do not make any attempt to further constrain the
dynamics associated with $\Sigma$.  This question has to be solved
experimentally by measuring the free parameters, looking for new
states related to $\Sigma$, and comparing this to any theoretical
predictions in specific models.

The \index{fermions, masses}fermion mass term in the effective
Lagrangian is replaced by
\begin{equation}\index{operators, dimension-3}
  \LL_3 = - (\bar Q_L\Sigma M_QQ_R + \bar L_L\Sigma  M_LL_R + \hc)
        - \bar L_R^c M_{N_R}\tfrac{1+\tau^3}{2} L_R
\end{equation}
which has the required $SU(2)_L\times U(1)_Y$ symmetry.

The boson mass term is replaced by a kinetic-energy term for the
$\Sigma$ field.  We introduce further abbreviations
\begin{gather}
  \index{$V_\mu$ field}
  V_\mu = \Sigma (D_\mu\Sigma)^\dagger \quad\text{and}\quad
  \index{$T$ field}
  T = \Sigma \tau^3 \Sigma^\dagger
  \\
  \text{where}\quad
  \index{covariant derivative}
  D_\mu\Sigma = \partial_\mu\Sigma + ig\vW_\mu\Sigma - ig'\Sigma\vB_\mu
\end{gather}
to write this as
\begin{equation}
  \LL_{2(W)} = -\tfrac{v^2}{4}\tr{V_\mu V^\mu}
\end{equation}
In unitary gauge, the vector field $V_\mu$ corresponds to a particular
combination of the $W$ and $B$ fields, namely
\begin{align}
  V_\mu &= - ig\vW_\mu\Sigma + ig'\Sigma\vB_\mu 
\end{align}
which defines the physical massive vector bosons, $W^\pm_\mu$ and
$Z_\mu$.

The dimension-four part of the Lagrangian does not involve the
$\Sigma$ field
\begin{equation}
\begin{split}
  \index{operators, dimension-4}
  \LL_4 &= \bar Q_L i\fmslash D Q_L + \bar Q_R i\fmslash D Q_R
   + \bar L_L i\fmslash D L_L + \bar L_R i\fmslash D L_R
  \\ &\quad
   - \tfrac12 \tr{\vW_{\mu\nu} \vW^{\mu\nu}} 
   - \tfrac12 \tr{\vB_{\mu\nu} \vB^{\mu\nu}},
\end{split}
\end{equation}
Allowing for \index{$\Sigma$ field}$\Sigma$ self-interactions
$\LL_\Sigma$, the complete \index{effective Lagrangian}effective
Lagrangian
\begin{equation}\label{chpt}
  \LL = \LL_{2(W)} + \LL_3 + \LL_4 + \LL_\Sigma
\end{equation}
is invariant under the full electroweak symmetry group.  In analogy
with low-energy \index{QCD}QCD, this effective theory is called the
\index{chiral Lagrangian}\emph{chiral Lagrangian} of electroweak
interactions.  Its validity is not restricted to the
particular scenario of dynamical symmetry breaking it is usually
associated with.  The minimal \index{Standard Model}SM with a Higgs
boson is just a special case of (\ref{chpt}), where the field $\Sigma$
is given a definite linear representation.

\subsection{Anomalous couplings}
\label{sec:anomalous}

The guideline for constructing the \index{effective
Lagrangian}effective Lagrangian~(\ref{chpt}) has been to start with
the \index{Fermi model}Fermi model Lagrangian and add the minimal
set of fields that make the weak-interaction symmetries manifest.
This requires the addition of kinetic terms for the new fields and the
inclusion of $\Sigma$ factors in the boson and fermion mass terms.
However, if one does \index{loop effects}one-loop calculations
with~(\ref{chpt}), proper \index{gauge fixing}gauge-fixing and
\index{ghost terms}ghost terms taken into
account, one will observe that additional operators are needed to make
the theory finite at next-to-leading order.  This is natural since the
Lagrangian~(\ref{chpt}) does not yet contain all possible
\index{operators, dimension-4}operators of
dimension four or less consistent with electroweak symmetry.

At dimension two, there is an additional operator not yet considered,
namely
\begin{equation}\label{rho-op}
  \index{operators, dimension-2}
  \LL_{2(W)}' = -\beta'\tfrac{v^2}{8}\tr{T V_\mu}\tr{T V^\mu}.
\end{equation}
Imposing CP-invariance on the effective Lagrangian\footnote{A
discussion of \index{CP violation}CP violation is beyond the scope of
this review.}, the complete list of dimension-four operators not
contained in~(\ref{chpt}) reads\cite{EWChPT}
\begin{align}
  \index{operators, dimension-4}
  \LL_1 &= \alpha_1 
         gg'\tr{\Sigma\vB_{\mu\nu}\Sigma^\dagger\vW^{\mu\nu}} \label{L1}\\
  \LL_2 &= i\alpha_2 
         g'\tr{\Sigma\vB_{\mu\nu}\Sigma^\dagger[V^\mu,V^\nu]} \label{L2}\\
  \LL_3 &= i\alpha_3 g\tr{\vW_{\mu\nu}[V^\mu,V^\nu]} \label{L3}\\
  \LL_4 &= \alpha_4(\tr{V_\mu V_\nu})^2 \label{L4}\\
  \LL_5 &= \alpha_5(\tr{V_\mu V^\mu})^2 \label{L5}\\
  \LL_6 &= \alpha_6 \tr{V_\mu V_\nu} \tr{TV^\mu} \tr{TV^\nu} \label{L6}\\
  \LL_7 &= \alpha_7 \tr{V_\mu V^\mu} \tr{TV_\nu} \tr{TV^\nu} \label{L7}\\
  \LL_8 &= {\tfrac14}\alpha_8 g^2 (\tr{T\vW_{\mu\nu}})^2\label{L8}\\
  \LL_9 &= {\tfrac{i}{2}}\alpha_9 g
		 \tr{T\vW_{\mu\nu}}\tr{T[V^\mu,V^\nu]} \label{L9}\\
  \LL_{10} &= \tfrac12\alpha_{10} (\tr{TV_\mu}\tr{TV_\nu})^2\label{L10}\\
  \LL_{11} &= \alpha_{11}g\epsilon^{\mu\nu\rho\lambda}
		\tr{TV_\mu}\tr{V_\nu \vW_{\rho\lambda}}\label{L11}
\end{align}

In the general case of a nonlinear symmetry representation the
Lagrangian contains terms of arbitrarily high
dimension\index{operators, higher dimension}.  Therefore, this list is
not sufficient to make the theory finite to \emph{all} orders.  In
each order of perturbation theory new terms are introduced with the
dimension of the $\Sigma$-dependent terms increased by two.

This fact does not make the \index{effective
Lagrangian}effective-Lagrangian approach useless.  It merely implies
that at each order of the perturbative expansion one should be
prepared for new contributions which are generically of the order
$1/16\pi^2$ (since they are induced as loop corrections) with the
operator dimension increased by two\cite{NDA}.  The two extra powers
of fields or derivatives are compensated by two powers of $1/v$, the
expansion parameter of $\Sigma$ in~(\ref{exp-par}).  As long as the
energy is small enough, one can truncate the perturbative series to
obtain an approximation of the true amplitude.  In matrix elements,
the \index{loop expansion}loop expansion therefore becomes a
\index{low-energy expansion}\emph{low-energy} expansion in terms of
\begin{equation}
  \frac{E^2}{(4\pi v)^2} = \frac{E^2}{\Lambda^2} .
\end{equation}
where $E$ is any linear combination of energies, masses and momenta
assigned to the external particles.  This sets the scale where
perturbation theory breaks down in the absence of Higgs-like states:
\begin{equation}
  \Lambda = 4\pi v \approx \q{3}{\TeV}.
\end{equation}
Of course, there may be larger contributions of the
operators~(\ref{L1}--\ref{L11}) than predicted by the loop expansion.
Since the effect of anomalous couplings increases with energy,
generically this leads to a lower \index{cutoff scale}cutoff scale
$\Lambda'<\Lambda$.

\subsection{Custodial symmetry}

\index{custodial symmetry}
As we have seen in the previous section, there are free parameters
which arise when massive vector bosons are introduced to regularize
the \index{Fermi model}Fermi model.  One of those affects the
$M_W/M_Z$ ratio which is usually referred to as the \index{$\rho$
parameter}$\rho$ parameter\cite{SU2c}:
\begin{equation}\label{rho}
  \frac{M_W^2}{M_Z^2 c_w^2} = \rho \quad\text{where}\quad
  \rho = \frac{1}{1+\beta'}.
\end{equation}
Experimentally, $\rho$ is approximately equal to unity.  Hence, the
coefficient $\beta'$ of the operator~(\ref{rho-op}) in the chiral
Lagrangian vanishes to leading order in perturbation theory.

This can be attributed to an approximate symmetry\cite{SU2c,EFT}.  If
we take the global symmetry of the Lagrangian not to be $SU(2)_L\times
U(1)_Y$ but enlarge it to $SU(2)_L\times SU(2)_R$ (where
$U(1)_Y\subset SU(2)_R$ ) and impose the transformation law on
\index{$\Sigma$ field}$\Sigma$
\begin{equation}
  \Sigma \to U\,\Sigma V^\dagger
\end{equation}
with $U\in SU(2)_L$ and $V\in SU(2)_R$, then the term~(\ref{rho-op})
is forbidden.  Simultaneously, this symmetry forbids all operators in
the chiral Lagrangian which contain a \index{$T$ field}$T$ factor in
the trace, namely $\LL_6$ to $\LL_{11}$ (\ref{L6}--\ref{L11}).

It is natural to take the \index{fermions, representation}right-handed
fermions to be doublets under $SU(2)_R$, just as the left-handed
fermions are doublets under $SU(2)_L$.  However, for the fermions this
is not a good symmetry.  It is violated by the up-down mass
differences (and by right-handed neutrino \index{Majorana
masses}Majorana masses).

The hypercharge part of weak interactions also violates this symmetry.
The $B$ vector field, in our notation, couples to $\Sigma$ by a
$\tau^3$ matrix factor, which is not consistent with a right-handed
$SU(2)_R$ symmetry.  Nevertheless, this breaking is proportional to
$s_w^2$ which is not a large parameter, and the fermion couplings
affect vector and scalar interactions at the loop level only.  Looking
at bosonic interactions, $SU(2)_R$ invariance is a reasonable
approximation.  

Spontaneous symmetry breaking in this sector then takes the form
\begin{equation}
  SU(2)_L \times SU(2)_R \to SU(2)_C
\end{equation}
where $SU(2)_C$ is the diagonal subgroup, the \index{custodial
symmetry}\emph{custodial $SU(2)$ symmetry} of electroweak
interactions.  Bosonic states should come as $SU(2)_C$ multiplets, and
in fact, the $Z$ and $W^\pm$ masses are degenerate up to the factor
$1/c_w^2\approx 1 + s_w^2$.  Similarly, one expects new particles
associated with EWSB also to be organized as $SU(2)_C$
multiplets.\footnote{This argument is independent of the origin of
EWSB; in Higgs models, the spectrum also tends to follow this
pattern.}

\section{Goldstone boson scattering}
\label{sec:Goldstone}

\index{Goldstone bosons}
In the chiral-Lagrangian approach described in this chapter,
electroweak symmetry breaking and fermion mass generation are mediated
by a matrix-valued field \index{$\Sigma$ field}$\Sigma$ which
parameterizes the \index{Higgs sector}Higgs sector dynamics.  The
degrees of freedom that make up $\Sigma$ are a probe for the mechanism
of electroweak symmetry breaking.  A complete theory would probably
replace $\Sigma$ by a multitude of new (elementary or composite)
fields, but in any case the scalars $w^a$ introduced in the minimal
parameterization~(\ref{exp-par}) must be present to serve as
\index{vector bosons, longitudinal} longitudinal $W$ and $Z$ bosons.
This is the triplet of Goldstone bosons associated with spontaneous
electroweak symmetry breaking\cite{Goldstone}.  Thus, the only
experiment which
\emph{guarantees} information about electroweak symmetry breaking is
a measurement of Goldstone boson interactions.

In the high-energy limit ($s,t,u$ all going to infinity) corresponding
scattering amplitudes of $W_L^\pm,Z_L$ on the one hand and $w^\pm,w^0$
on the other hand become equal.  This fact is known as the
\index{equivalence theorem}\emph{Equivalence Theorem}\cite{EQT}.  We
can make use of it by considering scattering amplitudes of Goldstone
bosons in place of the electroweak gauge bosons that are observed in
the detector (i.e., their decay products).  Doing this, we should
always keep in mind that for a quantitative analysis one has to
compute the full electroweak amplitudes, since realistic collider
energies are far from the asymptotic region.

\subsection{Quasielastic scattering at leading order}

The lowest-order effective Lagrangian~(\ref{chpt}) provides a unique
prediction for the quasielastic $2\to 2$ scattering amplitudes of $W$
and $Z$ bosons.  In the absence of $SU(2)_C$ violation, projecting
onto longitudinal states and taking the high-energy limit one obtains
the \index{low-energy theorem}\emph{Low-Energy Theorem} (LET)\cite{LET}
\begin{align}
  A(W^-_LW^-_L\to W^-_LW^-_L) &= -{s}/{v^2}\label{LET-wwww}\\
  A(W^+_LW^-_L\to W^+_LW^-_L) &= -{u}/{v^2}\\
  A(W^+_LW^-_L\to Z_LZ_L) &= {s}/{v^2} \\
  A(Z_LZ_L\to Z_LZ_L) &= 0 \label{LET-zzzz}
\end{align}
The cross sections for on-shell scattering are calculated by squaring
the amplitudes, inserting phase space factors and dividing by a
symmetry factor of two for like-sign $W$ and for $ZZ$ final states.
This symmetry factor will not be included in the amplitude in any of
the relations given here.

To be precise, the LET predicts the numerical coefficient of the
lowest-order term of an expansion of the scattering amplitudes in
terms of $E/v$, terms of order $M_W/E$ neglected.  Since $M_W=gv/2$,
this is in fact the limit $g\to 0$ with $v$ fixed.  As an
approximation to the exact amplitude, the LET is useful for energies
larger than $M_W$ and below the scale where either new states appear
or partial-wave unitarity is saturated otherwise (see
Sec.~\ref{sec:unitarity}).

\subsection{Custodial symmetry relations}

\index{custodial symmetry}\index{Goldstone bosons}
The Goldstone bosons transform under $SU(2)_C$ transformations as a
triplet.  Therefore, if this symmetry is exact, all quasielastic
scattering amplitudes are expressible in terms of a single function
\index{$A(s,t,u)$}$A(s,t,u)$:
\begin{align}
  A(W^-_LW^-_L\to W^-_LW^-_L) &= A(t,s,u) + A(u,t,s) \label{SU2c-wwww}\\
  A(W^+_LW^-_L\to W^+_LW^-_L) &= A(s,t,u) + A(t,s,u) \\
  A(W^+_LW^-_L\to Z_LZ_L) &= A(s,t,u) \\
  A(W^-_LZ_L\to W^-_LZ_L) &= A(t,s,u) \\
  A(Z_LZ_L\to Z_LZ_L) &= A(s,t,u) + A(t,s,u) + A(u,t,s)\label{SU2c-zzzz}
\end{align}
The function $A(s,t,u)$ satisfies
\begin{equation}
  A(s,u,t) = A(s,t,u).
\end{equation}
As we have seen, its Taylor expansion begins with
\begin{equation}
  A(s,t,u) = {s}/{v^2}.
\end{equation}
Note that these relations are strongly violated in forward scattering
where photon exchange is important.  There, elastic $WW$
scattering becomes singular while $WW\to ZZ$ stays finite.  This is
outside the validity region of the \index{equivalence
theorem}Equivalence Theorem.

\subsection{Next-to-leading order contributions}

\index{loop effects}
Even without additional knowledge about the high-energy behavior of
the theory, the next-to-leading order corrections to Goldstone
scattering can be computed.  Only the logarithmic terms are
scheme-independent and thus physically
meaningful\cite{DW89,NDA,CG87}:
\begin{equation}\label{A-1loop-log}
\begin{split}
  \Re A(s,t,u) = \frac{s}{v^2}
    &+ \frac{1}{16\pi^2v^4}\left\{
      -\frac{(t-u)}{6}\left[t\ln\frac{-t}{\mu^2}-u\ln\frac{-u}{\mu^2}\right]
      -\frac{s^2}{2}\ln\frac{s}{\mu^2}\right\}
      \\
    & + \alpha_4^0\frac{4(t^2+u^2)}{v^4} + \alpha_5^0\frac{8s^2}{v^4}.
\end{split}
\end{equation}
The result depends on a renormalization scale $\mu$.  This dependence
can be absorbed in a redefinition of the coefficients of the operators
$\LL_4$ and $\LL_5$:
\index{operators, dimension-4}
\begin{align}
  \alpha_4(\mu) &= \alpha_4^0 
    - \frac{1}{12}\,\frac{1}{16\pi^2}\ln\frac{\mu^2}{\mu_0^2}, &
  \alpha_5(\mu) &= \alpha_5^0
    - \frac{1}{24}\,\frac{1}{16\pi^2}\ln\frac{\mu^2}{\mu_0^2}
\end{align}
Finite corrections depend on the calculational scheme, i.e., on the UV
completion of the theory.  They are contained in the constant
coefficients $\alpha_{4,5}^0$, which therefore represent the relevant
information.  The same applies to the $SU(2)_C$-violating couplings
$\alpha_{5,6,10}$ which are scale-independent to this order, if the
coefficient $\beta'$ (\ref{rho-op}) is indeed zero (or $\rho=1$), as
suggested by data.

\subsection{Unitarity constraints}
\label{sec:unitarity}

\index{unitarity}
The optical theorem states that the total cross section for any
process is equal to the imaginary part of the elastic forward
scattering amplitude.  If there is only elastic $2\to 2$ scattering,
this can be translated into a relation for the scattering amplitude
$A(s,t,u)$.  Expanding it in \index{partial-wave amplitudes}partial
waves
\begin{equation}\index{$A(s,t,u)$}
\begin{split}
  A(s,t,u) &= 32\pi\sum_\ell a_\ell(s)\,(2\ell+1)\, P_\ell(1 + 2t/s),
\end{split}
\end{equation}
each partial-wave amplitude $a_\ell$ has to satisfy
\begin{equation}\label{Argand}
  |a_\ell(s) - i/2| = 1/2,  
\end{equation}
i.e., as a curve in the complex plane parameterized by $s$ it has to
stay on the \index{Argand circle}\emph{Argand circle}, a circle with
radius $\frac12$ around the point $\frac{i}{2}$.  In particular, the
real part of the partial-wave amplitude can never exceed $1/2$.

In terms of the amplitude function $A(s,t,u)$, the $SU(2)_C$
eigenamplitudes are given by\cite{Uni}
\begin{align}\index{$A(s,t,u)$}
  A(I=0) &= 3A(s,t,u) + A(t,s,u) + A(u,t,s), \\
  A(I=1) &= A(t,s,u) - A(u,t,s), \\
  A(I=2) &= A(t,s,u) + A(u,t,s).
\end{align}
Inserting the \index{low-energy theorem}LET expressions
(\ref{LET-wwww}--\ref{LET-zzzz}) and expanding in terms of
partial-wave amplitudes, one obtains the nonvanishing terms
\begin{align}
  a_{J=0}^{I=0} &= \frac{s}{16\pi v^2}, & 
  a_{J=1}^{I=1} &= \frac{s}{96\pi v^2}, &
  a_{J=0}^{I=2} &= -\frac{s}{32\pi v^2}. \label{a-LET}
\end{align}
There is no higher spin involved if we remain with the LET amplitudes.
The critical value $a=1/2$ is reached at the energies
\begin{align}
  I&=0: & E&=\sqrt{8\pi}\, v = \q{1.2}{\TeV}, \\
  I&=1: & E&=\sqrt{48\pi}\,v = \q{3.5}{\TeV}, \\
  I&=2: & E&=\sqrt{16\pi}\, v = \q{1.7}{\TeV}.
\end{align}

At these energies, perturbation theory ceases to be predictive.  In
order to have amplitude expressions that are at least in accord with
unitarity beyond these scales, one can try to resum the perturbation
series in a particular way.  The result depends on the chosen
resummation prescription and does not tell anything about the actual
high-energy behavior.  However, it can serve as a consistent
implementation of particular models with distinct features at energies
beyond the unitarity saturation threshold.

\index{unitarity}
The idea of such unitarization models is to project each
eigenamplitude function $a_\ell(s)$ onto the \index{Argand
circle}Argand circle.  Doing this, one assumes implicitly that no new
scattering channels are open, so that $2\to 2$ quasielastic scattering
dominates at all scales.  Two particular models have become popular,
representing extreme cases:
\begin{enumerate}
\item
\index{$K$-matrix unitarization}
The $K$-matrix unitarization model\cite{Kmatrix,Kmatrix2} is not
limited to the perturbative expansion.  Assuming that $a(s)$ is a
real-valued amplitude function one starts with, the unitarized
amplitude is given by
\begin{equation}\label{K-matrix}
  a_K(s) = a(s)\,\frac{1 + i a(s)}{1 + a(s)^2}.
\end{equation}
Geometrically, the value $a_K(s)$ corresponds to the projection of the
point $a(s)$ onto the Argand circle along the straight line connecting
$z=a(s)$ with $z=i$ (Fig.\ \ref{fig:K-matrix}).  By construction, the
$K$-matrix prescription will never generate a resonance if there is
none within the function $a(s)$.  In this respect it can be regarded
as a minimal unitarization model.  In particular, if the LET
expression $a(s)=s/v^2$ is inserted, the amplitude function $a(s)$
translates into
\begin{equation}
  a_K(s) = a_0 s\,\frac{v^2 + i a_0 s}{v^4 + a_0^2 s^2}.
\end{equation}
This unitarized amplitude will asymptotically approach the fixed point
$a_K(s)=i$, a resonance at infinity.

\begin{figure}
\begin{center}
\includegraphics{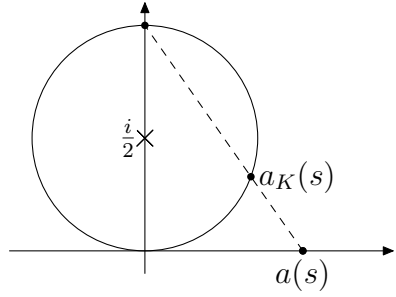}
\end{center}
\caption{$K$ matrix construction for projecting a real scattering
amplitude onto the Argand circle.}
\label{fig:K-matrix}
\end{figure}

\item
\index{Pad\'e unitarization}
To obtain the Pad\'e unitarization model (also known as the
\index{inverse amplitude method}inverse
amplitude method)\cite{Pade}, one separates the amplitude into two
pieces.  Usually, one takes the leading term $a^{(0)}(s)$ and the real
part of the next-to-leading order term $a^{(1)}(s)$ in the chiral
expansion which are proportional to $s$ and to $s^2$, respectively.
Then, the unitarized amplitude reads
\begin{equation}
  a_P(s) = \frac{a^{(0)}(s)^2}{a^{(0)}(s) - a^{(1)}(s) - ia^{(0)}(s)^2}
\end{equation}
If $a^{(1)}(s)$ vanishes, this coincides with the $K$-matrix model.
However, if $a(s)$ has the form
\begin{equation}
  a(s) = a_0\left(\frac{s}{v^2} + \alpha\frac{s^2}{v^4}\right)
\end{equation}
the Pad\'e-unitarized amplitude is
\begin{equation}
\begin{split}
  a_P(s) &= \frac{-a_0 s/\alpha}{s - v^2/\alpha + ia_0 s/\alpha}
\end{split}
\end{equation}
This is a \index{resonances in Goldstone scattering}resonance with
mass $M = {v}/{\sqrt{\alpha}}$ and width $\Gamma = {a_0M}/{\alpha}$.
In other words, adopting the Pad\'e unitarization method is equivalent
to the assumption that a $2\to 2$ scattering amplitude is entirely
dominated by a single resonance.
\end{enumerate}
Pad\'e unitarization works remarkably well for pion-pion scattering in
low-energy \index{QCD}QCD, consistent with vector meson dominance.
Unfortunately, this does not imply anything for the scattering of
electroweak Goldstone bosons.

\subsection{Resonances and new particles}
\label{sec:res}

\index{resonances in Goldstone scattering}\index{Goldstone bosons}
A striking signature of new physics is a resonance in some scattering
channel.  If such resonances appear in Goldstone boson scattering,
this would almost certainly give a clue in the search for the origin
of electroweak symmetry breaking.  While resonances are more likely to
be observed at the \index{LHC}LHC where the available energy for
Goldstone boson scattering is somewhat higher than for a first-stage
Linear Collider, the strongest resonance in a particular scattering
channel will have a low-energy tail that contributes to the parameters
in the low-energy expansion.  Resonances could be elementary particles
(such as a Higgs boson) or bound states of more
fundamental objects yet to be discovered.

As discussed before, one expects states associated with EWSB to be
grouped in multiplets of the $SU(2)_C$ \index{custodial
symmetry}custodial symmetry.  Given the $SU(2)_C$ quantum numbers of
Goldstone bosons, this leaves the following possibilities:
\begin{align}
  &\text{Scalar singlet $\sigma$:} 
    &\LL_{\sigma}&= g_{\sigma} \sigma \tfrac{v}{2}\tr{V_\mu V^\mu} 
  \label{s1}\\
  &\text{Vector triplet $\rho^a_\mu$:} 
    &\LL_{\rho}&= g_{\rho}\tfrac{v^2}{2}\tr{\rho^a_\mu\tau^a V^\mu}
  \label{v3} \\
  &\text{Tensor singlet $\tau^{\mu\nu}$:}
    &\LL_{\tau}&= g_{\tau}\tfrac{v}{2} \tau^{\mu\nu}\tr{V_\mu V_\nu}
  \label{t1} \\
  &\vdots&\vdots\nonumber
\end{align}
The amplitude functions corresponding to these couplings for the
scalar and vector cases take the form
\begin{align}\index{$A(s,t,u)$}
  A_{\sigma}(s,t,u) &= g_{\sigma}^2\frac{s^2}{v^4}\,\frac{1}{s-M^2} ,
  \label{A-s1} \\
  A_{\rho}(s,t,u) &= g_{\rho}^2
       \left(\frac{s-u}{t-M^2} + \frac{s-t}{u-M^2}
             + 3\frac{s}{M^2}\right).
  \label{A-v3}
\end{align}
where the resonance width has to be inserted when a pole falls
inside the physical region.

\index{resonances in Goldstone scattering}
\index{scalar resonance}\index{vector resonance}\index{tensor resonance}
While scalar and tensor resonances are electrically neutral, a vector
resonance multiplet $\rho$ has neutral and charged components
analogous to the $W^\pm,Z$ triplet.  The interaction resulting from
evaluating (\ref{v3}) is antisymmetric, forbidding the coupling of the
$\rho$ to identical particles: It can decay into $W^+W^-$ but not into
$ZZ$.

If $SU(2)_C$ is violated, resonances transforming as scalar triplets
($a_0$), vector singlets ($\omega$) etc.\ are also accessible in
Goldstone scattering, and the amplitude relations
(\ref{SU2c-wwww}--\ref{SU2c-zzzz}) between $Z$ and $W$ external
states are lost.

\index{scalar resonance}\index{$\sigma$ resonance}
A scalar resonance $\sigma$ with arbitrary coupling $g_\sigma$ is the
generalization of a \index{Higgs boson}Higgs resonance.  The width of
such a state is given by
\begin{equation}
  \Gamma_\sigma = \frac{3g_\sigma^2}{32\pi}M_\sigma.
\end{equation}
if Goldstone bosons are the only decay channels.  In the SM, the
coupling $g_\sigma$ itself is proportional to the mass, $g_\sigma =
\sqrt2\,M/v$, and thus $\Gamma\propto M^3$.  Such a state becomes very
broad and loses its identity if $M\gtrsim\q{1}{\TeV}$.  Beyond the SM,
the $\sigma$ mass and coupling need not be related, and narrow scalar
resonances may be present.

\index{resonances in Goldstone scattering}
\index{vector resonance}\index{$\rho$ resonance}
A vector resonance triplet is a characteristic feature of QCD-like
\index{technicolor}
technicolor models\cite{dynEWSB,BESS}.  If there are no other decay
channels than Goldstone bosons (i.e., longitudinal vector bosons), the
resonance width is given by
\begin{equation}
  \Gamma_\rho = \frac{g_\rho^2}{48\pi}M_\rho
\end{equation}
which is smaller than the width of a scalar with equal mass and
coupling.

Vector resonances have the special property that they can mix with
electroweak gauge bosons.  This may be interpreted as a remnant of the
electroweak interactions of their constituents.  As a result, one
expects a significant sensitivity to $\rho$ properties not just in
Goldstone scattering, but also in the $e^+e^-\to W^+_LW^-_L$
scattering amplitude.
%(Fig.~\ref{graphs:rho-mix}).

\index{resonances in Goldstone scattering}
\index{pseudoscalar resonance}
A special case are neutral pseudoscalar resonances, $\pi^0_T$ and
$\eta^0_T$.  While they do not couple to Goldstone pairs, they can have a
coupling to pairs of \emph{transversal} vector bosons which is induced
by the triangle anomaly: The coupling strength is much smaller than
for $\sigma$ and $\rho$ states, and one expects the coupling to
$ZZ$, $W^+W^-$, $\gamma\gamma$ and even $gg$ to be similar in
magnitude.  This is to be contrasted with $\sigma$ resonances for
which the (longitudinal) $ZZ$ and $WW$ couplings are dominant.
However, this picture may be complicated by CP violation in the strong
dynamics, which would induce $\eta-\sigma$ mixing.  Clearly, if any
resonance appears in vector boson scattering, it is important to
determine its spin and the polarization of vector bosons in the decay
by angular correlation analysis.

Apart from Goldstone (vector boson) decays, all such states are likely
to have a significant or even dominant fraction of heavy-quark decays:
\index{top quark}
$t\bar t$, $t\bar b$, $b\bar b$.  The channels $\tau^+\tau^-$ and
$\tau^+\nu_\tau$ are also possible.

\index{Goldstone bosons}
While resonant production of new states in Goldstone scattering is
restricted by the symmetries of Goldstone pairs, any state associated
with EWSB can in principle be pair-produced.  Some particles can also
be directly pair-produced in $e^+e^-$ annihilation, but Goldstone
scattering gives access to additional members of new multiplets.  The
coupling strength extracted from the production cross section of such
a particle is an independent piece of information.

\index{pseudo-Goldstone bosons}
In models of dynamical symmetry breaking, low-lying pseudo-Goldstone
boson scalars are a common feature\cite{dynEWSB,HS02}.  Classifying
them according to their $SU(2)_C$ properties, multiplets analogous to
the low-energy QCD spectrum can be expected, among them $SU(2)_C$
triplets ($\pi$), doublets ($K$) and singlets ($\eta$).  All can in
principle be produced in Goldstone scattering, and the cross sections
may be sizable.  As dominant decay modes one expects longitudinally
polarized vector bosons and heavy quarks, possibly accompanied by
transversally polarized vector bosons (radiative decays).

\index{technipions}
Pseudo-Goldstone bosons (technipions) share quantum numbers with the
$H^\pm,A^0$ states that are present in models with more than one Higgs
doublet, e.g., the MSSM.  As a consequence, the detection of
Higgs-like scalars is \emph{not} sufficient to establish a weakly
interacting scenario of EWSB.  Only by a careful analysis of the
complete pattern of masses and couplings a particular model can be
favored or excluded, and the measurement of the couplings to Goldstone
bosons (longitudinal $W,Z$ as opposed to transversal gauge bosons) is
an important ingredient.  Many of the corresponding measurements are
difficult or impossible at a hadron collider even if production rates
are large, but are straightforward in the low-background environment
of a Linear Collider.

\section{Measuring Higgs sector parameters at a Linear Collider}

\subsection{Precision observables}
\label{sec:precision}

\index{effective Lagrangian}
The effective Lagrangian~(\ref{chpt}) and~(\ref{rho-op}--\ref{L11})
encodes all known facts about the structure of electroweak
interactions.  Our current knowledge about the free parameters of this
effective theory can be summarized as follows:
\begin{enumerate}
\item The masses of charged fermions and vector bosons have been measured
or derived from hadronic data with high accuracy.
\item \index{neutrinos}
The neutrino mass matrices are much less certain. We know about
three different eigenstates with very low mass and considerable
mixing, but nothing about any other eigenstates.
\item \index{weak currents}
The current-current structure of weak interactions is well
established for the first two generations of quark and leptons.  For
the third generation, there is still some room for contributions that
do not fit in this picture.
\item \index{operators, dimension-5}
Dimension-five operators (magnetic moments, flavor-changing
penguin operators, etc.) are suppressed, and all measurements and
limits are consistent with loop effects of the known particles.
\end{enumerate}
In short, the gauge symmetry structure is extremely well tested, at
least for the first two fermion families.  This makes the
\index{chiral Lagrangian}chiral-Lagrangian approach a meaningful
parameterization.  Any anomalous effect can consistently be
parameterized by the coefficients of gauge-invariant
(higher-dimensional) operators.  In the bosonic sector in particular,
a basis of the CP-conserving \index{operators,
dimension-4}dimension-four operators is given
by~(\ref{rho-op}--\ref{L11}).  These terms are sensitive to
\index{Higgs sector}Higgs
sector physics since they contain factors of the symmetry-breaking
\index{$\Sigma$ field}$\Sigma$ field, and therefore carry the
information about EWSB that is available at low energies until new
degrees of freedom are observed directly.

Only three parameters in this list are significantly constrained by
the electroweak precision data gathered during the last decade.  In
the analysis of $Z$ pole observables, of the $W$ mass and of the
flavor-independent low-energy data, all deviations from the SM
prediction can, to leading nontrivial order, be parameterized by the
coefficients of the three operators that are bilinear in the vector
fields.  In our terminology, these are $\alpha_1$, $\beta'$ and
$\alpha_8$ (\ref{L1}, \ref{rho-op}, \ref{L8}).  Another
parameterization has been introduced by Peskin and
Takeuchi\cite{STU}, who considered the Taylor expansion of vector
boson propagators.  The leading deviations from the SM relations are
given by three parameters \index{$S,T,U$ parameters}$\Delta S$,
$\Delta T$ and $\Delta U$, which are related to the chiral Lagrangian
coefficients by
\begin{align}\label{STU-alpha}
  \Delta S &=  -16\pi\alpha_1, &
  \alpha\Delta T &= -\beta', &
  \Delta U  &= -16\pi\alpha_8.
\end{align}
$\Delta T$ parameterizes custodial $SU(2)_C$ violation in the Higgs
sector, analogous to the $\rho$ parameter which is related to it
(\ref{rho}) by\index{$\rho$ parameter}\index{custodial symmetry}
\begin{equation}
  \Delta\rho = \alpha\,\Delta T.
\end{equation}
$\Delta S$ describes anomalous mixing of weak and hypercharge bosons
and thus affects the measured value of the weak mixing angle, while
$\Delta U$ parameterizes $SU(2)_C$ violation in the left-handed gauge
sector.  In most models, there is little room for the latter effect,
so $\Delta U$ is usually not considered.

\index{loop effects}
It should be emphasized that, in the absence of a Higgs boson, the SM
radiative corrections to the interactions parameterized by $\Delta S$
and $\Delta T$ are logarithmically divergent.  To obtain numeric
values for them, one has to introduce a \index{cutoff scale}cutoff in
the low-energy effective theory.  It is customary to introduce a
(fictitious) \index{Higgs boson}Higgs boson for that purpose, so that
values for $\Delta S$ and $\Delta T$ have to be understood in
reference to some fixed Higgs boson mass.

\begin{figure}
\begin{center}
\scalebox{0.6}{{\includegraphics*{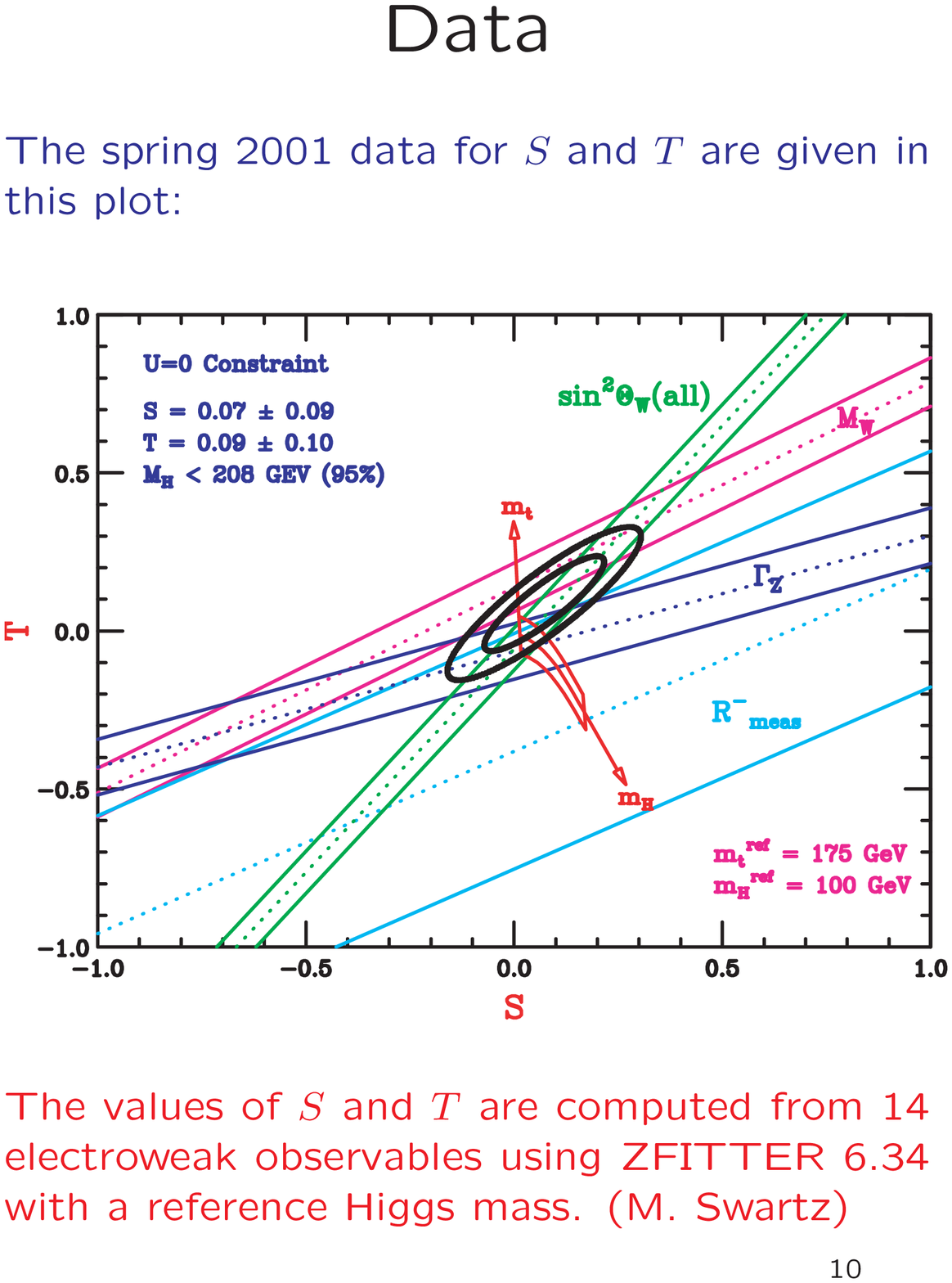}}}
\end{center}
\vskip-5mm
\caption{Exclusion countour in the $ST$ plane allowed by the
electroweak precision data.  The arrows denote the directions of
increasing top mass and increasing Higgs mass \protect\cite{ST01}.}
\label{fig:ST}
\end{figure}

\index{$S,T,U$ parameters}
The experimental constraints on $S$ and $T$ are summarized in
Fig.~\ref{fig:ST}, where the reference Higgs boson mass has been set
to $100\;\GeV$.  For this value, the exclusion contour encloses the
origin, and $\Delta S$ and $\Delta T$ are consistent with zero.
Incidentally, this is the prediction of the \index{Standard
Model}minimal SM with a light Higgs boson, which is therefore
consistent with the precision data.  If the reference Higgs mass $m_H$
is changed to higher values, the origin of the $ST$ plane moves into
the lower right direction, as indicated in the plot.  Then, the Higgs
sector has to contribute nonzero shifts $\Delta S$ and $\Delta T$.  In
fact, the actual location of the exclusion contour requires a positive
value of $\Delta T$ to make the data consistent with a high effective
Higgs mass.  In the absence of a light Higgs boson one should
therefore expect new physics to provide a certain amount of custodial
$SU(2)_C$ violation\cite{PW01}.

In models of dynamical symmetry breaking, $S$ and $T$ (or $\alpha_1$ and
$\beta'$) receive contributions from the \index{compositeness
scale}compositeness scale $\Lambda$ which are of the order
$\Delta\alpha_1 \sim {v^2}/{\Lambda^2}$.  If $\Lambda\sim 4\pi v$, the
natural upper limit for $\Lambda$ in the absence of a Higgs resonance,
this correction is of the same order as the shift from a light to a
heavy Higgs boson in Fig.~\ref{fig:ST}.  Of course, the sign and
precise value of the new contributions is model-dependent.  While
\index{technicolor}QCD-like technicolor models are disfavored by the
data since they typically have $\Delta S>0$ and $\Delta T\approx 0$
(conserved $SU(2)_C$), \index{topcolor}topcolor models, for instance,
predict custodial symmetry violation and positive $\Delta T$, which
makes them consistent with data for a larger range of physical Higgs
masses.

\index{$S,T,U$ parameters}
The bands in Fig.~\ref{fig:ST} indicate the areas in the $ST$ plane
allowed by individual observables which depend on $\Delta S$ and
$\Delta T$.  It is important that they all intersect in the same
region, so that a meaningful exclusion contour can be drawn.  If this
were not the case, one would have to include operators of dimension
six in the analysis.  The magnitude of their contribution is
parameterically of the same order as two-loop radiative corrections
which, however, are strongly scheme-dependent in the absence of a
physical Higgs boson.

\index{loop effects}
The consistency of the individual bands in the present plot indicates
that the inclusion of higher-order effects is not yet necessary, but
significant improvements in the experimental sensitivity (as expected
from a \index{Giga-Z}Giga-Z experiment, cf.\ \chMoenig) would
provide such a level of precision that a two-parameter analysis in
terms of $S$ and $T$ becomes obsolete.  This is obviously desirable in
the context of weakly interacting models where all observables are
computable to higher order in perturbation theory, and information on
additional parameters in the theory can be gained in this way.
However, since it is unlikely that for strong interactions a reliable
calculation of next-to-next-to-leading order effects is possible, in
the context of dynamical EWSB the experimental coverage of a larger
subset of the operator coefficients (\ref{L1}--\ref{L11}) will be more
valuable.

\subsection{Triple gauge couplings}

\index{triple gauge couplings}\index{operators, dimension-4}
In the operator basis (\ref{L1}--\ref{L11}), the operators $\LL_2$,
$\LL_3$, $\LL_9$ and $\LL_{11}$ do not contribute to vector boson
two-point functions, but modify the trilinear couplings of the photon
and of the $W$ and $Z$ bosons.  A standard parameterization of a
CP-conserving triple gauge boson vertex has been introduced
in\cite{TGC},
\begin{equation}
\begin{split}\label{TGC}
  \LL_{WWV} &= g_{WWV}\Big[
    ig_1^V V_\mu(W^{-\nu} W_{\mu\nu}^+ - W_{\mu\nu}^- W^{+\nu})
   +i\kappa_V W_\mu^- W_\nu^+ V^{\mu\nu}
  \\ &\qquad
   +i\frac{\lambda_V}{M_W^2}W_\mu^{-\nu}W_\nu^{+\rho} V_\rho^{\;\mu}
   +g_5^V\epsilon^{\mu\nu\rho\sigma}
       (W^-_\mu\partial_\rho W^+_\nu - \partial_\rho W^-_\mu
    W^+_\nu)V_{\sigma}
   \Big],
\end{split}
\end{equation}
where $V$ denotes the photon ($V=\gamma$ or $A_\mu$) and the $Z$ boson
interactions with prefactors $g_{WW\gamma}=e$ and $g_{WWZ}=gc_w$,
respectively.  If all anomalous operator coefficients vanish, we have
\begin{align}
  g_1^\gamma&=g_1^Z=\kappa_\gamma=\kappa_Z=1, \\
  g_5^\gamma&=g_5^Z=\lambda_\gamma=\lambda_Z=0.
\end{align}
Nonzero coefficients of the dimension-four operators contribute the
following shifts\cite{EWPrec}\footnote{There are also shifts due to
$\alpha_1$, $\beta'$ and $\alpha_8$.  They are constrained already now
by the existing data as discussed in the previous section.  These
contributions, together with the one-loop radiative corrections, have
to be included, but can be assumed to be known in a complete triple
gauge boson coupling analysis.}
\begin{align}
  \Delta\kappa_\gamma &= 
     g^2\alpha_2 + g^2\alpha_3 + g^2\alpha_9, \\
  \Delta\kappa_Z &=
     -g'{}^2\alpha_2 + g^2\alpha_3 + g^2\alpha_9, \\
  \Delta g_1^Z &= \tfrac{1}{c_w^2}g^2\alpha_3, \\
  \Delta g_5^Z &= \tfrac{1}{c_w^2}g^2\alpha_{11}, \\
  \Delta g_1^\gamma &= \Delta g_5^\gamma = 0, \\
  \Delta\lambda_\gamma &= \Delta\lambda_Z = 0.
\end{align}

\index{triple gauge couplings}
Due to electromagnetic gauge invariance, corrections to $g_1^\gamma$
and $g_5^\gamma$ have to vanish at zero momentum transfer.  However,
the absence of corrections to the $\lambda$ couplings up to this order
is a characteristic feature of the strongly-interacting scenario.
Nonzero values for these coefficients are only introduced at higher
order in the chiral expansion, i.e., by \index{operators,
dimension-6}dimension-six operators.  (In the weakly-interacting
scenario where a light Higgs boson is present, all anomalous terms
scale as dimension six.)  The reason is that $\lambda$ multiplies a
term that involves \index{vector bosons, transversal}transversal
vector fields only and thus does not probe the Higgs sector directly.
By contrast, the operators $\LL_2$, $\LL_3$, $\LL_9$ and $\LL_{11}$
(\ref{L2}, \ref{L3}, \ref{L9}, \ref{L11}) involve Goldstone scalars,
visible as the \index{vector bosons, longitudinal}longitudinal
components of vector bosons.  Hence, the strongly interacting scenario
predicts the transversal couplings $\lambda_\gamma$ and $\lambda_Z$ to
be significantly suppressed compared to possible deviations in the $g$
and $\kappa$ parameters.  Projecting onto longitudinal polarization
states of the vector bosons by exploiting angular correlations of
their decay products will enhance the relevant contributions.

\index{custodial symmetry}
With exact custodial $SU(2)_C$ symmetry we have
$\alpha_9=\alpha_{11}=0$ and get the additional relations
\begin{equation}
  \Delta\kappa_\gamma = -\tfrac{c_w^2}{s_w^2}(\Delta\kappa_Z-\Delta g_1^Z)
  \quad\text{and}\quad
  g_5^Z = 0,
\end{equation}
If this symmetry assumption is valid, the leading anomalous effect on
the couplings depends on just two parameters, $\alpha_2$ and
$\alpha_3$, which could be measured by considering $Z$ observables
only, $g_Z$ and $\kappa_Z$.

\index{triple gauge couplings}
The measurement of $W$ and $Z$ pair production at LEP2 has provided
the first meaningful bounds on anomalous triple gauge boson couplings.
The current precision is still low compared to that one already
achieved for $\Delta S$ and $\Delta T$.  This situation will change
when a Linear Collider is available.  As discussed in \chMoenig,
the experimental accuracy on the vector boson self-interactions will
then become competitive, so the indirect sensitivity to the Higgs
sector structure can be considerably improved by the complete coverage
of the operator basis.

\subsection{$W$ and $Z$ scattering amplitudes}
\label{sec:pheno}

\index{quartic gauge couplings}
The remaining parameters $\alpha_{4,5,6,7,10}$ (\ref{L4}--\ref{L7},
\ref{L10}) in the CP-conserving chiral Lagrangian do not contribute at
tree-level to bilinear or trilinear vector boson couplings.  Rather,
they affect Goldstone boson scattering and thus show up as anomalous
\emph{quartic} couplings of (longitudinally polarized) $W$ and $Z$
bosons.

\index{custodial symmetry}
From the list of operators introducing quartic couplings, only $\LL_4$
and $\LL_5$ conserve custodial $SU(2)_C$, so in the symmetric case the
quartic vector boson couplings depend on just two new parameters,
$\alpha_4$ and $\alpha_5$.  The other terms, $\LL_6$, $\LL_7$ and
$\LL_{10}$, describe new sources of $SU(2)_C$ violation.  This counting
does not include the effect of the parameters $\alpha_{3,8,9,11}$
which also contribute anomalous quartic vector-boson interactions.  As
discussed before, by combining low-energy and high-energy Linear
Collider data the latter will be sufficiently constrained to be
considered fixed in the analysis of quartic couplings, which must also
include the (calculable) one-loop SM corrections.

\index{quartic gauge couplings}\index{Goldstone bosons}
Concentrating on the genuine quartic couplings, in the $SU(2)_C$
symmetric limit the contribution of $\alpha_4$ and $\alpha_5$ to the
Goldstone scattering amplitude is given by~(\ref{A-1loop-log}).
However, this amplitude is not accessible directly, but is embedded in
multi-fermion processes involving intermediate gauge boson production
and decay.  There are three different classes of processes which probe
Goldstone scattering at lepton colliders:
\begin{enumerate}
\item
The one-loop amplitude for the production of a final state $VV'$ (where
$V,V'=W^\pm,Z$) is affected by a
\index{rescattering correction}\emph{rescattering correction}:
\begin{center}
\includefmgraph{dewsb-graphs}{1}{2mm}(50,15)
\end{center}

In this process rescattering of the vector bosons takes place at the
full c.m.\ energy of the annihilating fermions.  To make use of this
fact, in a global fit of the parameters describing pair production
their imaginary part has to be extracted.  Only the channel with spin
and isospin $1$ can be accessed.

\item
The second process class which is sensitive to the symmetry-breaking
sector is \index{triple vector boson production}\emph{triple vector
boson production}:
\begin{align}
  e^-e^+ &\to Z W^+ W^- \\
  e^-e^+ &\to Z Z Z
\end{align}
\begin{center}
\includefmgraph{dewsb-graphs}{2}{2mm}(45,15)
\end{center}

Gauge invariance makes the cross section for all processes of this
type fall off with $1/s$.  Therefore, the best measurements are not
necessarily done at the highest energy but somewhat above threshold,
and the sensitivity is limited by the available luminosity.  The
variable to project out resonances or to observe the effect of
anomalous couplings is the invariant mass of vector boson pairs.

\item
\index{Goldstone bosons}
The obvious place to look for Goldstone boson scattering amplitudes is
\index{vector boson fusion}\emph{vector boson fusion}:
\begin{align}
  e^-e^+ &\to e^-e^+ VV' \\
  e^-e^+ &\to \nu\bar\nu VV'
\end{align}
\begin{center}
\includefmgraph{dewsb-graphs}{3}{0mm}(45,25)
\vspace{2mm}
\end{center}
This class of processes has a cross section that rises logarithmically
with energy, so increasing the energy as well as the luminosity will
improve the experimental sensitivity.  In the asymptotic limit where
masses can be neglected, the intermediate vector bosons are
essentially on-shell and the relevant Goldstone scattering amplitudes
are probed directly, albeit at an effective c.m.\ energy which is
significantly lower than the full collider energy.
\end{enumerate}

Which process is actually most sensitive depends on machine parameters
and experimental details (cf.\ \chMoenig).  Concerning
\index{rescattering correction}rescattering corrections,
disentangling the imaginary parts of all form factors in vector boson
pair production (i.e., four-fermion production) is a nontrivial task.
Collider runs with different combinations of electron and positron
polarization are needed for a clean separation of all
contributions\cite{DNN02}.

The other processes involve six-fermion production in $e^+e^-$
collisions, where in the case of \index{triple vector boson
production}triple vector boson production all three fermion pairs
originate from vector boson decays, such that the signal can be
isolated by invariant mass constraints.  In the case of \index{vector
boson fusion}vector boson fusion, the spectator
\index{neutrinos}neutrinos (electron/positron) go predominantly into
the forward direction, hence the characteristic signature of this
process is a large missing invariant mass ($e^+e^-$ invariant mass).

\begin{figure}
\begin{center}
\scalebox{0.4}{{\includegraphics*{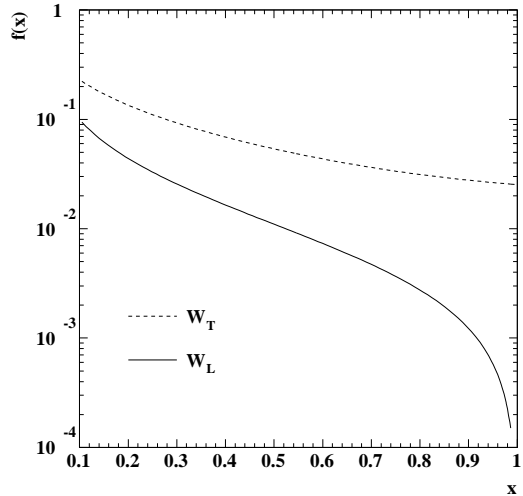}}}
\end{center}
\vskip-5mm
\caption{Structure functions for the emission of a transversally
and longitudinally polarized $W$ boson from an electron or
positron.}
\label{fig:EWA}
\end{figure}

For realistic Linear Collider parameters, triple vector boson
production\cite{WWZ} appears to be less sensitive to the parameters
of Goldstone scattering than vector boson fusion\cite{BHKPYZ} once
the collider energy is sufficient for the latter process to have a
significant rate.  In the asymptotic high-energy limit, this rate can
be approximated by the cross section for the on-shell subprocess of
$2\to 2$ vector boson scattering, folded by the splitting
probabilities $e\to W\nu$ (or $e\to Ze$).  \index{effective $W$
approximation}These structure functions are plotted in
Fig.~\ref{fig:EWA}.

The figure shows that the emission of \index{vector bosons,
longitudinal}longitudinally polarized $W$ bosons is significantly
suppressed compared to \index{vector bosons, transversal}transversally
polarized ones, in particular towards the high-energy end of the
spectrum.  Only the former probe the \index{quartic gauge
couplings}anomalous quartic interactions we are interested in, so
there is a significant $W_T$-induced background which has to be
reduced by a suitable experimental strategy\cite{BCHP,BHKPYZ,CR00}.
Since the longitudinal spectrum drops sharply near $x=1$, the energies
that can be reached for Goldstone boson scattering are considerably
lower than the collider energy.  For that reason, a Linear Collider
with an energy of at least $0.8$ to $1\;\TeV$ is necessary to achieve
a reasonable precision in the determination of the quartic couplings.
At these energies,
\index{unitarity}unitary constraints on the
amplitudes~(Sec.~\ref{sec:unitarity}) are not yet an issue, and the
chiral Lagrangian parameterization describes the scattering processes
in a model-independent way.

\index{quartic gauge couplings}
Anomalous quartic couplings involving photons will not be induced by
the dimension-four operators $\LL_4$ to $\LL_{10}$.  Therefore, in
$e\gamma$ and $\gamma\gamma$ collisions vector boson pair production
does not provide independent information on a strongly interacting
Higgs sector at this level, and in $e^+e^-$ collisions photon-induced
processes should be considered as a background to vector boson fusion.
Dimension-six operators, however, allow for anomalous quartic
couplings involving photons and can be probed independently in these
channels.

A complete coverage of the parameter space, which requires also the
inclusion of the $SU(2)_C$ violating operators in the analysis, will
be possible only by combining all available channels and including, in
particular, results for the analogous processes at the \index{LHC}LHC.
Nevertheless, the results presented in \chMoenig\ show that by
considering only the dominant vector boson fusion channels at a Linear
Collider, the experimental precision on the quartic couplings
$\alpha_4$ and $\alpha_5$ will be in the percent range, not much worse
that the expected accuracy in determining the bilinear and trilinear
couplings discussed before.\footnote{In comparing numerical values,
one should take into account that some authors extract a factor
$16\pi^2$ from the $\alpha$ parameters, such that their natural values
are~$\mathcal{O}(1)$.}

\section{Conclusions}

\index{dynamical symmetry breaking}
Dynamical symmetry breaking provides a natural explanation for the
electroweak scale.  If there is no weakly interacting effective theory
which describes physics beyond the $\TeV$ range, as it might be the
case if a light \index{Higgs boson}Higgs boson exists, the dynamics
responsible for electroweak symmetry breaking could be directly
accessible at future colliders.  \index{Goldstone bosons} The study of
Goldstone-boson scattering amplitudes is then the key for accessing
the (strongly-interacting) Higgs sector of the Standard Model.

The inherent scale of new strong interactions is likely beyond
$\q{1}{\TeV}$, therefore in this chapter we have considered mainly the
indirect effects on precision observables at a Linear Collider where
initially the energy may not be sufficient to produce new states.
However, compared to existing precision data the set of observables
that can be measured at a Linear Collider with at least percent
accuracy is greatly enlarged, and essentially all interactions in the
leading nontrivial order of the low-energy expansion are covered.
Together with hadron collider data, this will allow to significantly
constrain the possible scenarios and open the path towards a
satisfactory theory of electroweak symmetry breaking and mass
generation.

%%%%%%%%%%%%%%%%%%%%%%%%%%%%%%%%%%%%%%%%%%%%%%%%%%%%%%%%%%%%%%%%%%%%%%%%
%%% References for the chapter on dynamical EWSB
%%%%%%%%%%%%%%%%%%%%%%%%%%%%%%%%%%%%%%%%%%%%%%%%%%%%%%%%%%%%%%%%%%%%%%%%

\newcommand{\etal}{\textit{et al.}}

\clearpage

\printindex		%to generate and print out for index text  

\end{document}